\documentstyle[12pt]{article}

\begin{document}


\hsize=6.15in
\vsize=8.2in
\hoffset=-0.42in
\voffset=-0.3435in

\normalbaselineskip=24pt\normalbaselines

\begin{center}
{\large \bf Approximate invariance of metabolic energy per synapse during 
development in mammalian brains}
\end{center}

\vspace{0.15cm}

\begin{center}
{Jan Karbowski$^{*}$}
\end{center}

\vspace{0.05cm}

\begin{center}
{\it Institute of Biocybernetics and Biomedical Engineering, \\
Polish Academy of Sciences, Warsaw, Poland }
\end{center}


\vspace{0.1cm}

\begin{abstract}
During mammalian development the cerebral metabolic rate correlates 
qualitatively with synaptogenesis, and both often exhibit bimodal temporal
profiles. Despite these non-monotonic dependencies, it is found based on
empirical data for different mammals that regional metabolic rate per synapse 
is approximately conserved from birth to adulthood for a given species
(with a slight deviation from this constancy for human visual 
and temporal cortices during adolescence). A typical synapse uses about 
$(7\pm 2)\cdot 10^{3}$ glucose molecules per second in primate cerebral 
cortex, and about 5 times of that amount in cat and rat visual cortices.  
A theoretical model for brain metabolic expenditure is used to estimate 
synaptic signaling and neural spiking activity during development. 
It is found that synaptic efficacy is generally inversely correlated 
with average firing rate, and additionally, synapses consume a bulk of 
metabolic energy, roughly $50-90 \%$ during most of the developmental 
process (except human temporal cortex $ < 50\%$). 
Overall, these results suggest a tight regulation of brain 
electrical and chemical activities during the formation and consolidation 
of neural connections. This presumably reflects strong energetic 
constraints on brain development.
 
\end{abstract}





\vspace{0.1cm}

\noindent $^{*}$ Corresponding author at: jkarbowski@mimuw.edu.pl; 
jkarb@its.caltech.edu.

\vspace{0.3cm}

\newpage

{\Large \bf Introduction}

\vspace{0.3cm}

The proper functioning of neural circuits depends on their proper
wiring \cite{douglas,chechik,kaas,karb2001,karb2003,laughlin2003,chklovskii}. 
The right connectivity diagram is achieved during development that is both 
genetically and activity driven \cite{katz,cohen,lebe}, and which probably
has been optimized in the long evolutionary process \cite{allman,striedter}. 
Despite the widespread application of recording, imaging and molecular 
techniques \cite{silberberg,lendvai}, along with modeling studies 
\cite{chechik,karb2004,ooyen2003}, it is fair to say that our 
understanding of brain connectivity development is still very limited, 
and mostly qualitative. Nevertheless, the formation of neural circuits 
is an important problem in neuroscience, as its understanding may shed 
some light on structural memory formation in the brain and various 
developmental disorders \cite{rakic}. Moreover, synaptic development like 
every physical process requires some energy. A natural question is how much 
does it cost, and whether this cost changes during development. It is known 
that information processing in the brain is metabolically expensive 
\cite{aiello,levy,laughlin1998}. Specifically, energy consumption in 
mammalian brains increases fast with brain size, far more than in the 
rest of the body \cite{karb2007}.

The process of synaptogenesis, i.e. formation of synaptic connections, can 
be region specific and can have a complicated time-course, often bimodal 
with synaptic overproduction early in the development \cite{aghajanian,blue,
winfield,bourgeois1993,bourgeois1994,zecevic1989,zecevic1991,huttenlocher}. 
However, we do not know whether and how this process correlates with the 
activities of participating neurons. It is also unclear, to what extent the 
synaptogenesis is regulated metabolically, although some qualitative 
correlation between the two has been noted based on their temporal 
characteristics \cite{huttenlocher,chugani1998}.

A couple of theoretical investigations estimated that synapses in the adult
brain consume a significant portion of the overall metabolic rate 
\cite{attwell,lennie}. However, in fact, cerebral metabolic rate CMR (glucose
consumption rate) depends both on neural electric discharges and on
synaptic signaling, and their relative contribution is strongly controlled
by a neurotransmitter release probability and synaptic density \cite{karb2009}.
For instance, a high release probability can make synapses the major consumer
of energy, and conversely, a low probability can cause action potentials
to be metabolically dominant. Thus, simultaneous analysis of the cerebral 
metabolic rate and synaptic density during development can provide a useful 
quantitative information about the relative importance of these two factors. 
Additionally, it can yield a relationship between synaptic signaling and 
neural firing rates.

The main aim of this study is to address these questions in two steps. First,
by collecting and analyzing empirical data on brain metabolism and synaptic 
density during development for different mammals. Second, by combining 
these data with a theoretical model for brain metabolic rate \cite{karb2009},
in order to obtain quantitative results on the relationship energy vs.
synapses. In particular, we want to establish how common across mammals 
are mechanisms that relate synaptogenesis with neural activities and 
cerebral metabolism. A secondary goal is to test the analytic model of 
brain metabolism against the data, which is a little extended here from 
its original formulation in \cite{karb2009}. In this model, cerebral 
metabolic rate is expressed solely by neural and synaptic physiological 
parameters that are either known or can be easily measured.

\newpage

\noindent {\Large \bf Results}

\vspace{0.5cm}

\noindent {\bf Constancy of metabolic energy per synapse during development}

Empirical data (Tables 1-3) were used to 
analyze the time course of synaptic density ($\rho_{s}$) and glucose 
cerebral metabolic rate (CMR) during development for different mammals 
and brain regions (Fig. 1). For most regions both of these quantities 
depend non-monotonically on time, initially increasing, then reaching 
a maximum, and finally decreasing to adult values. In some cases, this
temporal dependence is even more irregular, with more than one maximum 
(e.g. rhesus monkey frontal cortex and human temporal cortex for synaptic 
density). Overall, CMR and $\rho_{s}$ can change several-fold during 
development. The most extreme change is in the cat visual cortex, where
$\rho_{s}$ and CMR can increase by a factor of $\sim$ 18 and $\sim$ 4,
respectively (Table 1). However, despite these complex dependencies and 
variability the amount of metabolic energy per synapse, i.e. the ratio 
CMR/$\rho_{s}$, is nearly independent of the developmental time for a 
given species and brain area (Fig. 2; Tables 1-3). In all examined
mammals and cortical regions, the quantity CMR/$\rho_{s}$ correlates weakly 
with the developmental time, and the linear slope in this dependence is 
close to zero. Moreover, these weak correlations are not statistically 
significant ($p$ value varies from 0.08 to 0.68; Fig. 2).

On average, rat 
brain consumes about $7\cdot 10^{-13}$ $\mu$mol of glucose per minute per 
synapse in the parietal cortex, and $(2-3)\cdot 10^{-12}$ $\mu$mol/min in 
the visual cortex (Table 1). The latter value is similar to the glucose 
use per synapse in the cat visual cortex (Table 1). In rhesus monkey and 
human cerebral cortices, there are approximately the same average baseline 
glucose consumptions per synapse, $\sim 7\cdot 10^{-13}$ $\mu$mol/min 
(Tables 2 and 3). From these results it follows that glucose use per synapse 
is smaller in large primate brains than it is in relatively small rodent 
of feline brains, and the difference could be five- or six-fold.

The biggest deviations from a baseline value of CMR/$\rho_{s}$ are for the 
human visual and temporal cortices between postnatal ages 3.5 and 12-15 years, 
and can be 2-3 folds above that baseline (Table 3). These numbers, however,
do not seem to be relatively large, considering that CMR in that period
can increase by a factor of 4-9 in relation to the minimal CMR. Nevertheless,
the ``energy per synapse'' distinction for the (pre- and) adolescent human 
brain is noticeable and could suggest a different distribution of energy 
in the developing human neural circuits in that period in comparison to 
other mammals.

\vspace{0.5cm}

\noindent {\bf Correlation between cerebral metabolic rate and synaptic density}

Empirical data on CMR and $\rho_{s}$ were used to find their mutual
relationship (Fig. 3). This relationship is in general monotonic with 
high positive correlations, and can be fitted by the formula, which was 
derived in the Materials and Methods:

\begin{equation}
\mbox{CMR}= a_{0} + [a_{1} + b\overline{\rho}_{s}]f(\overline{\rho}_{s}),
\end{equation}\\
where $a_{0}$ and $a_{1}$ are numerical coefficients that depend on
neurophysiological parameters (they are known and determined in the 
Materials and Methods), $b$ is the parameter related to synaptic
signaling, $\overline{\rho}_{s}$ is the amplitude of 
synaptic density, i.e. $\rho_{s}= \overline{\rho}_{s} 10^{11}$ [cm$^{-3}$].
The function $f(\overline{\rho}_{s})$ is the population average neural firing 
rate that changes during development with synaptic density as 
$f(\overline{\rho}_{s})= f_{0}\overline{\rho}_{s}^{c}$.
Values of the parameters $b$, $f_{0}$, and $c$ are determined by a fitting
procedure to the data, and they are presented in Table 4.

Generally, estimated average firing rates are rather small for all examined
mammals, and on average about 1 Hz (Table 4). The smallest values are 
for the monkey visual and sensorimotor cortices, and the largest for the cat 
visual cortex. The character of the relationship between population firing 
rate $f$ and synaptic density $\rho_{s}$ is not universal, but depends on 
a particular species and cortical region (Table 4). For some regions, the best 
fit is obtained for $f$ independent of $\rho_{s}$ (i.e. with $c\approx 0$). 
For others, we find an increase of $f$ with increasing $\rho_{s}$, 
either sublinearly ($ c < 1$) or approximately linearly ($c\approx 1$). 
The nature of this dependence has also its influence on the relationship 
CMR vs. $\rho_{s}$. When $c > 0$, that is, when $f$ increases with 
$\rho_{s}$, we find that CMR increases with $\rho_{s}$ in a non-linear manner 
(Fig 3B,C,D), whereas when $c= 0$, then CMR grows linearly with $\rho_{s}$ 
(Fig. 3A). Thus, we conclude that the dependence CMR on $\rho_{s}$ is
also non-universal.

\vspace{0.5cm}

\noindent {\bf Synaptic contribution to the cerebral metabolic rate 
during development}

Having determined the parameters $b$, $f_{0}$, and $c$, we can find
a fraction of metabolic energy consumed by synaptic signaling during
the development process. The fraction $\eta$ of the cerebral metabolic
rate CMR taken by synapses is defined as 
$\eta= b{\overline\rho}_{s}f({\overline\rho}_{s})/\mbox{CMR}$, or equivalently

\begin{equation}
\eta=  \frac{b f}{\mbox{CMR}/\overline{\rho}_{s}}.
\end{equation}\\
The latter expression implies that $\eta$ is inversely related to the
metabolic energy per synapse. Indeed, although $\eta$ changes during the
development much more than CMR/$\overline{\rho}_{s}$ (Tables 1-3), both of
these variables are negatively correlated (Table 5). The greater variability
of $\eta$ than CMR/$\overline{\rho}_{s}$ can be explained by its
additional dependence on firing rate $f(\overline{\rho}_{s})$, which in
itself is proportional to a variable synaptic density.

In general, $\eta$ is rather high, mostly in the range $0.5-0.9$ (Tables 1-3;
some $\eta$ is a little above unity, which is an artifact caused by systematic
errors in the fitting procedure that determines $b$, $c$, and $f_{0}$). 
A significant exception is human temporal cortex in which synapses use for
the most time considerably less than $50 \%$ of cortical CMR. At the top 
of the synaptogenesis, when synaptic density is maximal, $\eta$ is usually 
very large and often around 0.8-0.9, which is greater than for the adult,
but the difference is mild. From all examined mammals and cortical regions, 
synapses in the monkey visual and sensorimotor cortices, as well as synapses 
in the rat parietal cortex seem to be the most ``energetic'', since they 
frequently use approximately 90$\%$ of the total cerebral glucose rate.

Overall, these results strongly suggest that excitatory synaptic signaling 
uses a majority of metabolic energy allocated to neurons, even at
adulthood. The spiking neural activity and maintenance of negative membrane 
potential utilize generally far less energy, together approximately 
$10-40 \%$, depending on the species, brain region, and developmental period.

\vspace{0.5cm}

\noindent {\bf Relationship between synaptic efficacy and average firing 
rate across mammals}

The parameter $b$ in Eq. (1) is proportional to the excitatory synaptic
efficacy (or signaling; see Materials and Methods). For a given species,
we can associate this parameter with the average firing rate $f$, both of 
which were determined by fitting the theoretical model (Eq. 1) to the data 
(Table 4). We find that $b$ and $f$ are inversely correlated across all 
examined cerebral regions and animals, and can be fitted quite well by 
a universal curve of the form ($R^{2}= 0.956$; Fig. 4):

\begin{equation}
b= 0.03 f^{-1.35},
\end{equation}\\
where $b$ is expressed in $\mu$mol$\cdot$sec/min.
This relationship indicates that average synaptic efficacy is dependent
on network spiking activity, and the higher that activity the smaller 
synaptic signaling. For example, for $f=0.1$ Hz we have $b\approx 0.7$, while 
for $f= 1.3$ Hz we obtain $b\approx 0.03$, i.e. more than twenty-fold reduced 
synaptic efficacy. This implies that synaptic transmission is very sensitive
on the average firing rate in the network, which can have functional 
consequences (see Discussion).

\vspace{0.5cm}

\noindent {\bf Estimation of neurotransmitter release probability by
combining data and metabolic model}

Experimental data show that the probability of neurotransmitter release
is the least stable parameter among synaptic parameters, and
can change during the development by at least an order of magnitude
\cite{bolshakov,frick}. To test our metabolic model (see Materials and 
Methods), the release probability is estimated below for adult rat and
cat visual cortices. In this respect, we equate the empirical value of 
the parameter $b$ in Table 4 with the analytical formula for $b$ given 
by Eq. (15), which allows us to determine the release probability $q$. 
We assume that $\overline{g}_{AMPA}/\overline{g}_{NMDA} \approx 2.5$, in 
agreement with the empirical data for adult primate brain \cite{gonzalez}.
We take the peak AMPA synaptic conductances and their decay
time constants as: $\overline{g}_{AMPA}= 3.6\cdot 10^{-10}$ $\Omega^{-1}$
and $\tau_{AMPA}= 5\cdot 10^{-3}$ s for rat, and 
 $\overline{g}_{AMPA}= 7.1\cdot 10^{-10}$ $\Omega^{-1}$
and $\tau_{AMPA}= 7.6\cdot 10^{-3}$ s for cat \cite{yoshimura}. Additionally,
the NMDA synaptic conductance decay time constant $\tau_{NMDA}$ is
taken as $\tau_{NMDA}= 0.1$ s for both species, as a standard NMDA decay 
time \cite{jahr}. We find that the neurotransmitter release probability $q$
is 0.45 for adult rat visual cortex, and 0.31 for adult cat visual cortex.
These values are in the range of values reported experimentally
\cite{bolshakov,volgushev,murthy}, and suggest that the metabolic model 
presented and used in this paper (Materials and Methods) is reliable and 
has a predictive power.

\newpage

{\Large \bf Discussion}

This study shows that despite temporal changes in cerebral metabolic rate 
CMR and synaptic density $\rho_{s}$ during development, often exhibiting
bimodal shape, the amount of metabolic energy per synapse (CMR/$\rho_{s}$) 
is almost invariant in the process for a given mammal and brain region 
(Fig. 2; Tables 1-3). This approximate constancy is even more pronounced if 
we take into account that many other neuroanatomical parameters, such as 
neuron number, dendritic tree length, and brain volume, all change 
non-monotonically with an animal age \cite{pakkenberg,sowell,herndon}. 
In contrast to CMR/$\rho_{s}$, the fraction of CMR consumed by synapses,
i.e. $\eta$, is much more variable during the development (Tables 1-3).
Moreover, these two quantities are strongly negatively correlated (Table 5).
For the most developmental time and cortical regions $\eta$ is greater
than 0.5, implying that synapses use the majority of cortical metabolic
energy, often close to 90$\%$ or more (Tables 1-3).

The case with the human brain is more subtle, as its visual and temporal 
cortices exhibit a noticeable deviation from the CMR/$\rho_{s}$ constancy 
during early and middle adolescence (by a factor of $\sim$ 2; Table 3). 
In addition, $\eta$ for human temporal cortex is considerably smaller
than 0.5 for the most time. The increase in CMR/$\rho_{s}$ for the above
regions during adolescence is associated with a simultaneous decrease
in $\eta$, which suggests that non-synaptic part of CMR dominates over
the synaptic part in this period (Table 3). It is interesting to note
that the maxima of CMR/$\rho_{s}$ for human visual and temporal cortices
between 3.5 and 12 years coincide with maxima observed in cortical
volume, thickness, and surface area during the same time 
\cite{giedd,shaw,raznahan}. This positive (negative) correlation between
CMR/$\rho_{s}$ ($\eta$) and structural cortical growth can be an indication
that the latter process requires an additional energy above some baseline,
which is partly generated by shunting it from the synapses.

On average, a synapse in the primate cerebral cortex consumes about
$(7\pm 2)\cdot 10^{-13}$ $\mu$mol of glucose per minute. In rat and cat visual
cortices corresponding numbers are about 5 times larger, which qualitatively
agrees with a previous rough estimate that in larger brains energy per
synapse should be smaller than in smaller brains \cite{karb2007}.
These numbers translate into $5000-9000$ of consumed glucose molecules and
$(1.6-2.8)\cdot 10^{5}$ of consumed ATP molecules, both per second and per 
synapse in the primate cortex (using Avogadro number $\approx 6\cdot 10^{23}$ 
mol$^{-1}$, and the fact that about 31 ATP molecules are produces per one 
used glucose molecule \cite{rolfe}). Thus, the cost of creating and 
maintaining one synapse in the human cortex during development is about 
$2.2\cdot 10^{5}$ ATP molecules/second, which can increase during adolescence 
to $4\cdot 10^{5}$ ATP/sec.

There is a growing evidence that a typical excitatory synapse can operate
only in a limited number of structurally different discrete states 
\cite{luscher,montgomery}. Since the sizes of synapses (lengths of 
postsynaptic densities) during postnatal development remain roughly 
constant \cite{blue,zecevic1991}, one can assume that the number of synaptic 
states is also approximately invariant. Assuming that a synapse has on 
average between 10 and 100 states \cite{montgomery}, we can estimate the
amount of ATP utilization per 1 bit of stored synaptic information. For
human brain we obtain 
$2.2\cdot 10^{5}/\log_{2}10^{n} \sim (3-6)\cdot 10^{4}$ ATP/bit per second,
where $n= 1$ or 2. Thus, during a human lifetime ($\sim$ 80 years)
a typical synapse uses $(3\pm 1)\cdot 10^{14}$ ATP molecules per stored
1 bit of information.

Invariants in the brain design or dynamics are not too numerous, and their 
existence clearly deserves more attention and thought. The current finding 
about the constant energy per synapse during development (for a given brain 
region) expands a short list of the discovered invariants, including
adult synaptic density across mammals \cite{braitenberg,defelipe}, 
volume-specific metabolic scaling exponent across gray matter 
($\approx - 0.15$) \cite{karb2007}, energy per neuron across mammals 
\cite{houzel2011,karb2011}, blood flow and capillary length per neuron 
\cite{karb2011}, or fraction of brain volume taken by glia across 
mammals \cite{houzel2006,houzel2007}. It seems that there are some common 
principles underlying these invariants, which could be related to the 
economy of brain wiring 
\cite{kaas,karb2001,karb2003,mitchison,cherniak,wen,kaiser}.
This in turn could be associated with the evolutionary constraints 
coming from limited energetic resources \cite{levy,laughlin1998,attwell}, 
as the brain is an energy-expensive organ \cite{aiello,karb2007}, and synapses 
were pointed out as one of the important users of the cerebral metabolism 
\cite{karb2007,attwell,lennie,karb2009}. 
The fact that cerebral metabolic rate CMR and synaptic density $\rho_{s}$ 
are rather strongly positively correlated (Table 4, Fig. 3) speaks in 
support of the last argument.

The results in this study indicate that synapses are even bigger energy 
users than previously estimated. Calculations presented in Tables 1-3 show 
that at adulthood, when synaptic density is generally lower than in 
adolescence, synapses can still consume about $50-80 \%$ of the total 
glucose consumption rate. For example, for rat cortex $\eta$ is either
0.48 (visual) or 0.81 (parietal). The average of these values is about 
twice the amount that was previously calculated for adult rat cortex 
\cite{attwell}. The likely source of the discrepancy is the probability 
of neurotransmitter release, which was calculated here as 0.45 (for rat 
visual cortex), and assumed in \cite{attwell} as 0.25. Generally, it should 
be kept in mind that the computed values of the release probability are 
only averages, as this parameter is highly variable in time and additionally 
input specific, and could be somewhere between $0.05-0.7$ 
\cite{bolshakov,frick,volgushev,murthy}. Because the neurophysiological 
model of the gray matter metabolism presented in this paper (see Materials 
and Methods) yields reasonable numerical values of this highly uncertain 
parameter, it could play a useful role in the future in determining other 
functional circuit parameters from glucose metabolic data.

It is found that, as a rule, synaptic efficacy (signaling) is
negatively correlated with cortical average neural firing rate across
all examined species (Fig. 4). Low firing rates usually correspond to high 
synaptic efficacy, and vice versa (Fig. 4). The interesting feature is that 
all data points coming from different mammals and cortical regions collapse
(with high correlations) 
into one universal curve given by Eq. (3). This clearly suggests that 
synaptic regulatory mechanisms such as depression and potentiation are 
coupled with global network activity and may have a universal cross-species 
character. This kind of synaptic plasticity is reminiscent of the so-called 
synaptic scaling, which was found in cortical circuits \cite{turrigiano1998}.
In this process, which is typically slow, synaptic efficacy increases
if network activity is too low, and it decreases if network activity is too 
high. This synapse-network activity coupling serves as a tuning mechanism
to balance brain spiking activity, which may be important for preventing 
pathological dynamic states \cite{turrigiano2004}.

The collected empirical data in combination with the theoretical metabolic
model allow us to determine average firing rates across mammals during
development, from the birth to adulthood. These rates are rather low,
generally in the range $0-2.3$ Hz. This probably implies that only a small 
fraction of cells is active concurrently, which is compatible with
an idea of sparse neural coding in cortical networks \cite{levy,attwell}.
Moreover, our results show that larger brains tend to have a slightly lower 
spiking activities than smaller brains (Table 4). This conclusion that was 
reached here for developing brains is in line with a previous estimate made 
for several adult mammals, also using glucose metabolic data \cite{karb2009}. 
The current interesting finding is that neural firing rate could change during 
development in coordination with the changes in synaptic density (Table 4). 
Such dependence improves the goodness of fits for several brain regions 
significantly.

The semi-empirical results of this study can have some impact on modeling
studies related to the connectivity development in the brain. It has been
known for a long time that synaptic development is driven to some extent 
by global spiking activity of neurons \cite{lendvai,zito}. This coupling
has also been incorporated in several formal models dealing with 
synaptogenesis \cite{ooyen2003,ooyen1994}, but it often had abstract forms. 
It seems that the semi-empirical formula derived here (Eq. 3), allows us 
for a more realistic approach. Alternatively, this formula could be used 
as a one of the criterions for verification of modeling studies.
Similarly, the finding that there exist a (roughly) constant amount of 
available energy per synapse during development (Fig. 2; Tables 1-3), 
has not been explored in computational models. Yet, it could have important 
theoretical implications.

Although, the empirical data in this paper are concerned with normal 
development, they could also have some relevance for studies dealing with
developmental disorders, such as schizophrenia or autism. There are some
strong experimental indications that these mental diseases are associated
with altered synaptic connectivity \cite{mcglashan,geschwind}. It would
be interesting to know whether in these disorders the amount of metabolic
energy per synapse during development is also conserved or not? If not,
then how large are deviations form a constancy, and whether this measure
is somehow correlated with the degree of mental disorder. This perhaps could
have some practical applications.

\newpage

\noindent {\Large \bf Materials and Methods}

\vspace{0.3cm}

\noindent {\bf Developmental data}

The ethics statement does not apply to this study.
Experimental data for glucose cerebral metabolic rate (CMR) and synaptic 
density ($\rho_{s}$) during development for rat, cat, macaque monkey, and
human are presented in Tables 1-3. These mammals have adult brains that 
span 3 orders of magnitude in volume. The metabolic data were collected
from the following sources: for rat \cite{nehlig}; for cat \cite{chugani1991}; 
for monkey \cite{moore,jacobs,noda}; for human \cite{chugani1998,kinnala}. 
The synaptogenesis data were taken from: \cite{aghajanian,blue} for rat; 
\cite{winfield} for cat; 
\cite{bourgeois1993,bourgeois1994,zecevic1989,zecevic1991} for
monkey; and \cite{huttenlocher} for human.

\vspace{0.6cm}

\noindent {\bf Theoretical model of cerebral metabolic rate}

In this section we derive an expression for the glucose cerebral metabolic 
rate CMR in gray matter. This derivation follows closely a detailed analysis
presented in \cite{karb2009}, and additionally extends it by
including also NMDA synaptic currents. We assume that the activities of
Na$^{+}$/K$^{+}$ pumps are the major contributors to brain 
metabolism, which is in agreement with empirical estimates 
\cite{erecinska,ames}.
The main objective of these pumps is to remove Na$^{+}$ ions from
neuron's interior, in order to maintain a negative membrane resting
potential, which is critical for all neural functions.

During one cycle, the Na$^{+}$/K$^{+}$ pump extrudes 3 Na$^{+}$ and
intrudes 2 K$^{+}$ ions, which translates into a net removal of one
elementary positive charge that comprises a pump current $I_{p}$.
Consequently, the pump current $I_{p}$ constitutes of only 1/3 of the 
total sodium current through the membrane. In terms of the metabolic
cost, this pumping process uses 1 ATP molecule (per one cycle) to 
remove one positive charge. The metabolic expenditure of this process 
in the long run depends on the level of intracellular sodium 
concentration.

According to biochemical estimates \cite{rolfe}, about 31 ATP molecules 
are made per one oxidized glucose molecule during cellular respiration.
Consequently, the glucose metabolic rate CMR (the amount of moles of
glucose per tissue volume and time) is given by

\begin{eqnarray}
\mbox{CMR}= \frac{N \overline{I_{p}}}{31UF},
\end{eqnarray}\\
where $\overline{I_{p}}$ is the average net pump current, $N$ is the 
number of neurons contained in the gray matter volume $U$, and $F$ is
the Faraday constant. The ratio $\overline{I_{p}}/F$ is the amount of
moles of ATP molecules consumed on average per neuron per time unit.

At the steady state, i.e. for constant firing rates and after averaging
over long times (hundred of seconds to several minutes), the average
sodium concentration inside neurons is relatively stable \cite{karb2009}.
This corresponds to the situation when the pump current $\overline{I_{p}}$
balances 3 different types of sodium currents through the membrane 
\cite{karb2009}:

\begin{eqnarray}
3\overline{I_{p}}= I_{Na,0} + I_{ap} + I_{s,0},
\end{eqnarray}\\
where $3\overline{I_{p}}$ is the amount of Na$^{+}$ charge per second
that is removed by the Na$^{+}$/K$^{+}$ pump. The current $I_{Na,0}$ is 
Na$^{+}$ influx through sodium channels at rest (a small contribution), 
$I_{ap}$ is Na$^{+}$ influx due to action potentials, and $I_{s,0}$ is 
the sodium influx through synapses during background dendritic synaptic 
activity. The explicit forms of the first two currents
are given by:

\begin{eqnarray}
I_{Na,0}= g_{Na,0}S(V_{Na}-V_{0}),
\end{eqnarray}\\
\begin{eqnarray}
I_{ap}= f\overline{C}S(V_{Na}-V_{0}),
\end{eqnarray}\\
where $V_{Na}$ is the reversal potential for Na$^{+}$ ions, $V_{0}$ is 
the resting membrane potential, $f$ is the average firing rate, 
$g_{Na,0}$ is the resting Na$^{+}$ conductance per unit area,
$\overline{C}$ is effective membrane capacitance per unit area, and 
$S$ is the neuron's membrane surface area.

The synaptic contribution $I_{s,0}$ to the sodium influx is proportional
to a temporal average over an interspike interval of the AMPA and NMDA
synaptic currents, and takes the form: 

\begin{eqnarray}
I_{s,0}= \alpha M fq \int_{0}^{1/f} dt
\left[g_{AMPA}(t)+ G(V)g_{NMDA}(t)\right] V,
\end{eqnarray}\\
where $\alpha$ is the proportionality factor between the total synaptic 
current and Na$^{+}$ influx current and is given by 
$\alpha= V_{K}(V_{Na}-V_{0})/[V_{0}(V_{K}-V_{Na})]$, where $V_{K}$
is the reversal potential for K$^{+}$ ions. The latter dependence 
can be easily computed \cite{karb2009} and follows from the fact that AMPA 
current is composed exclusively of Na$^{+}$ and K$^{+}$ ions, and NMDA 
current is composed largely of these ions (the influence of Ca$^{+2}$ 
is neglected here, as it constitutes only of about 7-10$\%$ of the NMDA 
current \cite{burnashev}). The symbol $M$ denotes number of synapses per 
neuron, $q$ is the neurotransmitter release probability, and $V$ is neuron's 
membrane voltage. The function $G(V)$ is a voltage-dependent factor 
associated with NMDA receptors given by \cite{jahr}: 
$G(V)= 1/[1+0.33\exp(-0.06V)]$, where $V$ is in mV. For voltage equal to 
the resting potential, i.e. $V= V_{0}= -65$ mV, we obtain 
$G_{0}\equiv G(V_{0})= 0.06$. The symbols $g_{AMPA}(t)$ and $g_{NMDA}(t)$
denote the time dependent single synapse conductances, respectively AMPA 
and NMDA type. Below, we assume that the rising phase of these conductances
is much faster than their decaying phases. That is, we take
$g_{AMPA}= \overline{g}_{AMPA}\exp(-t/\tau_{AMPA})$, and
$g_{NMDA}= \overline{g}_{NMDA}\exp(-t/\tau_{NMDA})$, where 
$\overline{g}_{AMPA}$, $\overline{g}_{NMDA}$ are the peak conductances,
and $\tau_{AMPA}, \tau_{NMDA}$ are corresponding decay time constants.
Also, since the duration of a single action potential is very short
in comparison to the average interspike interval $1/f$, we can assume
that for the most time $V\approx V_{0}$ under the integral. With these 
assumptions we can carry out the integration in Eq. (8), with the result

\begin{eqnarray}
I_{s,0}= \frac{qfM V_{K}(V_{Na}-V_{0})}
{(V_{K}-V_{Na})}
\left[\overline{g}_{AMPA}\tau_{AMPA}R_{AMPA}(f)+
G_{0}\overline{g}_{NMDA}\tau_{NMDA}R_{NMDA}(f)
 \right],
\end{eqnarray}\\
where the frequency dependent factor $R_{i}(f)$ ($i= AMPA,NMDA$)
has the form: $R_{i}(f)= 1-\exp[-1/(f\tau_{i})]$. This factor for the
AMPA current is practically always close to 1, as $f\tau_{AMPA}$ is
significantly smaller than unity even for firing rates $f$ as large as
100 Hz (with $\tau_{AMPA} \sim 5-6$ msec). Generally, for the NMDA current
$R_{NMDA}$ is less than 1, and could be even $\ll 1$ for very large $f$.
However, for the empirical frequencies found in this study ($\sim 1$ Hz),
the factor $R_{NMDA}\approx 1$. Consequently, the values of $R_{AMPA}$
and $R_{NMDA}$ are both taken as 1 further in the analysis.

Combination of Eqs. (4-7) and (9) yields an approximate 
glucose metabolic rate CMR as follows:

\begin{eqnarray}
\mbox{CMR}= \frac{NS}{U}\frac{(V_{Na}-V_{0})}{93F}\left[ g_{Na,0}
+f\overline{C} + \frac{M}{S}\frac{V_{K}qf}{(V_{K}-V_{Na})}
(\overline{g}_{AMPA}\tau_{AMPA}+G_{0}\overline{g}_{NMDA}\tau_{NMDA})
\right].
\end{eqnarray}\\
Additionally, we assume that the geometry of
axons and dendrites can be approximated as cylindrical with equal volumes
\cite{braitenberg}. Thus, we can write the total membrane surface area
as $NS= 4(1-\phi)U/d$, where $d$ is an effective fiber diameter 
(harmonic mean of axonal and dendritic diameters), and $(1-\phi)$ is
the fraction of volume taken by neural wiring  \cite{karb2009}.
Moreover, the surface density of synapses can be written as
$M/S= \rho_{s}d/(4(1-\phi))$, where $\rho_{s}$ is the synaptic density
\cite{karb2009}. Substituting the above expressions for $NS/U$ and $M/S$
into Eq. (10), we obtain CMR in a more convenient form:

\begin{eqnarray}
\mbox{CMR}= \frac{(V_{Na}-V_{0})}{93F}\left[ \frac{4(1-\phi)}{d} 
\left(g_{Na,0} + f\overline{C}\right) + \frac{qf\rho_{s}V_{K}}
{(V_{K}-V_{Na})}(\overline{g}_{AMPA}\tau_{AMPA}
+ G_{0}\overline{g}_{NMDA}\tau_{NMDA}) \right],
\end{eqnarray}\\
or equivalently with an explicit dependence of CMR on synaptic density
and firing rate as:

\begin{eqnarray}
\mbox{CMR}= a_{0} + a_{1}f + b\overline{\rho}_{s}f,
\end{eqnarray}\\
where the coefficients $a_{0}$, $a_{1}$, and $b$ are given by
\begin{eqnarray}
a_{0}= \frac{4(1-\phi)g_{Na,0}(V_{Na}-V_{0})}{93Fd} ,
\end{eqnarray}\\
\begin{eqnarray}
a_{1}= \frac{4(1-\phi)\overline{C}(V_{Na}-V_{0})}{93Fd} ,
\end{eqnarray}\\
and
\begin{eqnarray}
b= \frac{10^{11}qV_{K}(V_{Na}-V_{0})
\left[\overline{g}_{AMPA}\tau_{AMPA}+ G_{0}\overline{g}_{NMDA}\tau_{NMDA}
\right]}{93F(V_{K}-V_{Na})}.
\end{eqnarray}\\
In Eq. (12) the firing rate $f$ is in Hz, and the symbol 
$\overline{\rho}_{s}$  denotes the synaptic density amplitude defined as 
$\rho_{s}= \overline{\rho}_{s} 10^{11}$, where $\rho_{s}$ 
is expressed in cm$^{-3}$. 
The coefficients $a_{0}$ and $a_{1}$ are invariant or nearly invariant
across species, and they do not seem to change significantly during
development after birth. This is because they depend on the parameters,
which themselves are developmentally or species independent. These are 
electrical voltages ($V_{Na}$, $V_{K}$, $V_{0}$) due to their logarithmic 
dependencies on ionic concentrations, membrane capacity $\overline{C}$, and 
structural parameters: the fraction of volume taken by wiring $(1-\phi)$ 
or fraction of neuropil \cite{bourgeois1994,zecevic1989,zecevic1991},
and the effective wire thickness $d$ \cite{braitenberg}. Also the sodium 
conductance at neuron's rest is very small, and biophysical models suggest 
that it is similar across species. 
The numerical values of these parameters are:
$V_{Na}= 0.050$ V, $V_{K}= -0.100$ V, $V_{0}= -0.065$ V (standard values), 
$(1-\phi)\approx 0.65$ 
\cite{bourgeois1994,zecevic1989,zecevic1991,braitenberg},
$g_{Na,0}= 3\cdot 10^{-7}$ ($\Omega$cm$^{2}$)$^{-1}$ \cite{karb2009},
$\overline{C}\approx 3.2\cdot 10^{-6}$ F/cm$^{2}$, and 
$d= 0.45\cdot 10^{-4}$ cm \cite{karb2009}. 
Based on these values, we obtain $a_{0}=0.013$ 
$\mu$mol/(g$\cdot$min), and $a_{1}=0.14$ $\mu$mol$\cdot$s/(g$\cdot$min).
The parameter $b$ is related to synaptic activities, and its value is 
determined in the Results section for every species and brain region.

There are no data on {\it in vivo} firing rates during development.
Therefore, we have to assume some form of $f$. We consider two scenarios 
for this quantity. In the simplest case, firing rate and synaptic density 
are independent of each other, and we take $f$ to be a constant. In a second 
case, we assume that firing rate and synaptic density are correlated in such 
a way that $f$ is an increasing function of $\overline{\rho}_{s}$. This 
follows from a simple expectation that higher synaptic density generally mean 
more excitatory synaptic input to a typical neuron, as $85 \%$ of synapses 
in the cerebral cortex are excitatory \cite{braitenberg,defelipe}. More 
excitatory input in a recurrent network translates into higher average firing 
rates. This is in agreement with mean-field models of recurrent neural networks 
\cite{brunel}. Thus, the simplest expression for the firing that combines
both scenarios is $f= f_{0}\overline{\rho}_{s}^{c}$, where $f_{0}$ and the 
exponent $c$ are to be determined by a fitting procedure to the data. 
When $c= 0$, then $f$ is independent of synaptic density.

\vspace{1.3cm}

%
%

\newpage

\vspace{1.5cm}



\newpage

{\bf \large Figure Captions}

Fig. 1\\
Dependence of glucose cerebral metabolic rate CMR and synaptic density 
$\rho_{s}$ on developmental time in visual cortex of various mammals. 
(A) Rat; (B) Cat; (C) Monkey; (D) Human. Circles correspond to 
the synaptic density and triangles to CMR.

\vspace{0.3cm}

Fig. 2\\
Approximate invariance of glucose cerebral metabolic rate per synapse 
during development. The linear fits to the data points are given in 
the brackets below. (A) Rat (circles - parietal cortex: 
$y=-0.0006x+0.092$, $R^{2}=0.259$, $p= 0.381$; squares - visual cortex:  
$y=0.0025x+0.187$, $R^{2}=0.223$, $p= 0.345$).
(B) Cat visual cortex (with the data point at 1 day: 
$y=-0.0023x+0.684$, $R^{2}=0.118$, $p= 0.451$; without the data point at
1 day: $y=0.0005x+0.322$, $R^{2}=0.438$, $p= 0.152$). 
(C) Monkey (circles - frontal cortex:
$y=0.00003x+0.063$, $R^{2}=0.279$, $p= 0.363$; 
squares - visual cortex:
$y=0.00009x+0.052$, $R^{2}=0.527$, $p= 0.102$; 
triangles - sensorimotor cortex:
$y=0.0007x+0.063$, $R^{2}=0.075$, $p= 0.656$).   
(D) Human (circles - frontal cortex:
$y=-0.0001x+0.116$, $R^{2}=0.025$, $p= 0.684$;
squares - visual cortex: 
$y=0.0002x+0.071$, $R^{2}=0.223$, $p= 0.282$;
triangles - temporal cortex:
$y=0.0003x+0.053$, $R^{2}=0.375$, $p= 0.079$). 
In the above fits $y$ refers to CMR/$\rho_{s}$ (in $10^{-11} \mu$mol/min) 
and $x$ to the developmental time (either in days for rat and cat or
in months for monkey and human). Note that for all fits the linear
coefficient is close to zero.

\newpage

Fig. 3\\
Empirical dependence of cerebral metabolic rate CMR on synaptic density
$\rho_{s}$ together with fits to the theoretical metabolic model. 
(A) Rat, parietal cortex. (B) Cat, visual cortex. (C) Monkey, visual 
cortex. (D) Human, frontal cortex. Empirical data are represented 
by diamonds, and theoretical fits by solid lines.
The fitting parameters are shown in Table 4. 

\vspace{0.3cm}

Fig. 4\\
Inverse relationship between synaptic signaling and average firing rate 
across mammals. Values of the synaptic efficacy $b$ and firing rates 
$f$ (arithmetic means) were found by fitting experimental data to the 
theoretical model (Table 4). Note that all data points (diamonds) coming 
from different species and cortical regions align into a universal curve 
of the form: $b= 0.03 f^{-1.35}$ ($R^{2}= 0.956$, $p < 0.001$).

\newpage

\begin{table}
\begin{center}
\caption{Synaptic and metabolic development for rat and cat cerebral cortex.}
\begin{tabular}{|l l l l l l|}
\hline
\hline
Species/region &  developmental  & $\rho_{s}$ & CMR 
&  CMR/$\rho_{s}$  &   $\eta$  \\
   &  time  &  [$10^{11}$ cm$^{-3}$]  &  [$\frac{\mu mol}{g\cdot min}$] 
&  [$10^{-11} \frac{\mu mol}{min}$]  &    \\

\hline
\hline

Rat:  &     &     &    &   &    \\
 $\;$  parietal cortex  &  14 day   &  2.8 $\;\cite{aghajanian}$  &  0.30 $\;\cite{nehlig}$  &  0.107 
  &   0.52    \\ 
    &  17 day   &  6.3 $\;\cite{aghajanian}$  &  0.42 $\;\cite{nehlig}$  &  0.067  &   0.84  \\
    &  21 day   &  9.0 $\;\cite{aghajanian}$  &  0.66  $\;\cite{nehlig}$ &  0.073  &   0.77  \\  
    &  35 day   &  14.0 $\;\cite{aghajanian}$  &  0.85 $\;\cite{nehlig}$  & 0.061  &   0.92  \\ 
    &  adult    &  13.5 $\;\cite{aghajanian}$  &  0.94 $\;\cite{nehlig}$  &  0.070  &  0.81  \\ 
                    &     &     &    &   &   \\
Rat:   &     &     &    &    &    \\
 $\;$  visual cortex   &  10 day   &  0.62 $\;\cite{blue}$  &  0.20 $\;\cite{nehlig}$  &  0.323
  &  0.10  \\ 
    &  14 day   &  1.16 $\;\cite{blue}$  &  0.24 $\;\cite{nehlig}$  &  0.207   &  0.29  \\
    &  17 day   &  2.68 $\;\cite{blue}$  &  0.32 $\;\cite{nehlig}$  &  0.119   &  1.19  \\  
    &  21 day   &  2.80 $\;\cite{blue}$  &  0.63 $\;\cite{nehlig}$  &  0.225   &  0.66  \\ 
    &  35 day   &  3.00 $\;\cite{blue}$  &  0.87 $\;\cite{nehlig}$  &  0.290   &  0.55  \\ 
    &  adult    &  2.95 $\;\cite{blue}$  &  0.97 $\;\cite{nehlig}$  &  0.329   &  0.48  \\ 
                    &     &     &    &    &  \\
Cat:   &     &     &    &    &     \\
 $\;$  visual cortex  &  1 day &  0.20 $\;\cite{winfield}$ &  0.318 $\;\cite{chugani1991}$ &  1.590 
  &   0.08  \\ 
    &  7 day &  0.50 $\;\cite{winfield}$ &  0.187 $\;\cite{chugani1991}$ &  0.374  &  0.42  \\
    & 30 day (est) & 2.50 $\;\cite{winfield}$ & 0.696 $\;\cite{chugani1991}$ & 0.278  &  0.89  \\  
    & 40-45 day  &  3.10 $\;\cite{winfield}$ & 0.987 $\;\cite{chugani1991}$ & 0.318   &  0.83  \\ 
    & 60-70 day  &  3.70 $\;\cite{winfield}$ & 1.406 $\;\cite{chugani1991}$ & 0.380   &  0.73  \\
    & 110-120 day & 3.10 $\;\cite{winfield}$ & 1.201 $\;\cite{chugani1991}$ & 0.387   &  0.68  \\ 
    &  adult    &  2.70 $\;\cite{winfield}$ & 1.120 $\;\cite{chugani1991}$ & 0.415    &  0.61  \\ 

\hline

\hline
\end{tabular}
\end{center}
Developmental time refers to postnatal time. References in the brackets.
Synaptic contribution $\eta$ to CMR is computed from Eq. (2).
\end{table}

\newpage

\begin{table}
\begin{center}
\caption{Synaptic and metabolic development for monkey cerebral cortex.}
\begin{tabular}{|l l l l l l|}
\hline
\hline
Species/region &  developmental  & $\rho_{s}$ & CMR 
&  CMR/$\rho_{s}$  &   $\eta$   \\
   &  time  &  [$10^{11}$ cm$^{-3}$]  &  [$\frac{\mu mol}{g\cdot min}$] 
&  [$10^{-11} \frac{\mu mol}{min}$]   &      \\

\hline
\hline

Monkey:  &     &     &    &     &      \\
 $\;$ frontal cortex   &  2-3 month   &  6.0 $\;\cite{bourgeois1994}$   &  0.33 $\;\cite{moore}$  &  0.055  
   &   0.63   \\
       &  4-5 month   &  6.1 $\;\cite{bourgeois1994}$  &  0.40 $\;\cite{moore}$ &  0.066  &  0.53  \\  
       &  6-7 month   &  5.7 $\;\cite{bourgeois1994}$  &  0.39 $\;\cite{moore}$ &  0.068  &  0.49  \\ 
       &  6 year   &  5.0 $\;\cite{bourgeois1994}$  &  0.34 $\;\cite{noda}$  &  0.068  &  0.47  \\  
       &  20 y (adult)  &  3.16 $\;\cite{bourgeois1994}$  &  0.22 $\;\cite{noda}$  &  0.070  &  0.36 \\ 
              &      &     &     &    &      \\
Monkey:   &     &     &    &    &     \\
 $\;$ visual cortex   &  0-2 month   &  5.5 $\;\cite{bourgeois1993}$  &  0.21 $\;\cite{jacobs}$  &  0.038 
  &  1.08   \\
       &  2-6 month   &  9.0 $\;\cite{bourgeois1993}$  &  0.50 $\;\cite{jacobs}$ &  0.056  &  0.94    \\
       &  8-9 month   &  8.0 $\;\cite{bourgeois1993}$  &  0.46 $\;\cite{moore}$ &  0.058   &  0.86   \\
       &  12 month    &  6.0 $\;\cite{bourgeois1993}$  &  0.33 $\;\cite{moore}$ &  0.055   &  0.78   \\ 
       &  6-7 year   &   6.0 $\;\cite{bourgeois1993}$   &  0.40 $\;\cite{noda}$  &  0.067  &  0.65   \\  
       &  20 y (adult)  &  3.8 $\;\cite{bourgeois1993}$   &  0.27 $\;\cite{noda}$  &  0.071  &  0.49  \\ 
               &     &     &     &    &      \\
Monkey:  &     &     &    &    &     \\
 $\;$ sensorimotor crtx  &  0-2 month   &  4.78 $\;\cite{zecevic1989,zecevic1991}$  &  0.26 $\;\cite{jacobs}$ 
& 0.054  &   1.20   \\
       &  2-3 month   &  5.75 $\;\cite{zecevic1989,zecevic1991}$  &  0.34 $\;\cite{moore}$ &  0.059  & 1.11  \\
       &  4-5 month   &  5.44 $\;\cite{zecevic1989,zecevic1991}$  &  0.44 $\;\cite{moore}$ &  0.081  & 0.81  \\
       &  6-7 month   &  5.19 $\;\cite{zecevic1989,zecevic1991}$  &  0.38 $\;\cite{moore}$ &  0.073  & 0.89  \\
       &  12-13 month  &  5.78 $\;\cite{zecevic1989,zecevic1991}$  &  0.37 $\;\cite{moore}$ &  0.064  & 1.03  \\

\hline

\hline
\end{tabular}
\end{center}
Developmental time refers to postnatal time. References in the brackets.
Synaptic densities for sensorimotor cortex are arithmetic means of values in motor
and somatosensory cortices.
\end{table}

\newpage

\begin{table}
\begin{center}
\caption{Synaptic and metabolic development for human cerebral cortex.}
\begin{tabular}{|l l l l l l|}
\hline
\hline
Species/region &  developmental  & $\rho_{s}$ & CMR 
&  CMR/$\rho_{s}$  &   $\eta$   \\
   &  time  &  [$10^{11}$ cm$^{-3}$]  &  [$\frac{\mu mol}{g\cdot min}$] 
&  [$10^{-11} \frac{\mu mol}{min}$]   &     \\

\hline
\hline

Human: &     &     &    &    &   \\
 $\;$ frontal cortex  & - (10-8) wbb(*) &  0.22 $\;\cite{huttenlocher}$   &  0.07 $\;\cite{kinnala}$ &  0.318 
  & 0.005 \\ 
       &  1 day       &  1.95 $\;\cite{huttenlocher}$ &  0.13 $\;\cite{kinnala,chugani1998}$ & 0.067  &  0.33  \\
       &  40 day      &  1.12 $\;\cite{huttenlocher}$ &  0.08 $\;\cite{kinnala}$ &  0.071  &   0.16 \\  
       &  80-83 day   &  3.10 $\;\cite{huttenlocher}$ &  0.15 $\;\cite{kinnala}$ &  0.048  &   0.81 \\ 
       &  1.17 year   &  3.79 $\;\cite{huttenlocher}$  &  0.26 $\;\cite{chugani1998}$ &  0.069 &  0.74 \\
       &  3.5 year    &  5.24 $\;\cite{huttenlocher}$  &  0.56 $\;\cite{chugani1998}$ &  0.107 &  0.70 \\
       &  12 year     &  4.69 $\;\cite{huttenlocher}$  &  0.44 $\;\cite{chugani1998}$ &  0.093 &  0.70 \\
       &  15 year     &  4.00 $\;\cite{huttenlocher}$  &  0.41 $\;\cite{chugani1998}$ &  0.103 &  0.53 \\  
       &  adult       &  3.40 $\;\cite{huttenlocher}$  &  0.27 $\;\cite{chugani1998}$ &  0.079 &  0.56  \\ 
               &     &     &     &    &      \\
Human:  &     &     &    &     &     \\
 $\;$ visual cortex    & - (10-8) wbb(*) &  1.2 $\;\cite{huttenlocher}$   &  0.06 $\;\cite{kinnala}$ &  0.050 
  &  0.98 \\ 
       &  1 day      &  2.6 $\;\cite{huttenlocher}$  &  0.18 $\;\cite{chugani1998}$  &  0.069  &  0.71  \\
       &  1 year     &  5.5 $\;\cite{huttenlocher}$  &  0.28 $\;\cite{chugani1998}$ &  0.051   &  0.96 \\
       &  1.5 year   &  4.9 $\;\cite{huttenlocher}$  &  0.32 $\;\cite{chugani1998}$ &  0.065   &  0.75 \\
       &  3.5 year   &  4.7 $\;\cite{huttenlocher}$  &  0.60 $\;\cite{chugani1998}$ &  0.128   &  0.38 \\
       &  12 year    &  3.6 $\;\cite{huttenlocher}$  &  0.45 $\;\cite{chugani1998}$ &  0.125   &  0.39 \\
       &  adult      &  3.1 $\;\cite{huttenlocher}$  &  0.27 $\;\cite{chugani1998}$ &  0.087   &  0.56 \\ 
                &     &     &     &    &      \\
Human:  &     &    &    &    &      \\
 $\;$ temporal cortex   & - (10-8) wbb(*) &  0.75 $\;\cite{huttenlocher}$  &  0.06 $\;\cite{kinnala}$  &  0.080
  &   0.06  \\ 
       &  1 day       &  2.94 $\;\cite{huttenlocher}$ &  0.09 $\;\cite{kinnala}$ &  0.031  &  0.41  \\
       &  40 day      &  2.10 $\;\cite{huttenlocher}$ &  0.07 $\;\cite{kinnala}$ &  0.033  &  0.30  \\  
       &  80-83 day   &  4.70 $\;\cite{huttenlocher}$ &  0.16 $\;\cite{kinnala}$ &  0.034  &  0.51  \\ 
       &  1.17 year   &  5.30 $\;\cite{huttenlocher}$ &  0.24 $\;\cite{chugani1998}$ &  0.045  &  0.42 \\
       &  3.5 year    &  5.57 $\;\cite{huttenlocher}$ &  0.52 $\;\cite{chugani1998}$ &  0.093  &  0.21  \\
       &  12 year     &  2.47 $\;\cite{huttenlocher}$ &  0.39 $\;\cite{chugani1998}$ &  0.158  &  0.07  \\
       &  15 year     &  3.89 $\;\cite{huttenlocher}$ &  0.36 $\;\cite{chugani1998}$ &  0.093  &  0.17  \\  
       &  adult       &  2.90 $\;\cite{huttenlocher}$ &  0.24 $\;\cite{chugani1998}$ &  0.083  &  0.15  \\

\hline

\hline
\end{tabular}
\end{center}
(*) Negative value refers to the weeks before birth (wbb). Positive
developmental times refer to postnatal time. References in the brackets.
\end{table}

\newpage

\begin{table}
\begin{center}
\caption{Best fits to the data for parameters in the relation CMR vs. $\rho_{s}$ 
across mammals.}
\begin{tabular}{|l l l l l l l|}
\hline
\hline
Species/region &  $b$ ($\mu$mol$\cdot$s/min)  & $c$  &  $f_{0}$ (Hz) 
& $f$ (Hz) &  R$^{2}$  &  SSE \\

\hline

Rat: parietal cortex  &  0.066 & 0.0  & 0.85  &  0.85  & 0.961  & 0.012  \\ 
Rat: visual cortex    &  0.071  & 1.02 & 0.73 & 0.4-2.2  & 0.674  & 0.181    \\ 
Cat: visual cortex    &  0.121  & 0.29 &  1.57 &  1.0-2.3  &  0.905  & 0.121  \\
Monkey: frontal cortex  &  0.024 &  0.52 &  0.57 & 1.0-1.5  & 0.776  & 0.011  \\
Monkey: visual cortex  &  0.228  & 0.48  & 0.08  &  0.15-0.23  & 0.908 & 0.005  \\
Monkey: sensorimotor crtx  &  0.692 & 0.03 & 0.09 &  0.1 & 0.262 & 0.013  \\
Human: frontal cortex   & 0.070  & 1.23  &  0.14 &  0.02-1.1  & 0.928 & 0.018  \\
Human: visual cortex   & 0.038  & 0.0  &  1.29  & 1.3  &  0.105 & 0.127   \\
Human: temporal cortex  & 0.010  &  0.69  &  0.60 &  0.5-2.0 &  0.347  & 0.142  \\

\hline

\hline
\end{tabular}
\end{center}

\end{table}

\newpage

\begin{table}
\begin{center}
\caption{Correlation between metabolic energy per synapse (CMR/$\rho_{s}$) and
synaptic fraction of metabolism ($\eta$).}
\begin{tabular}{|l l l |}
\hline
\hline
Species/region &  correlation  &  significance   \\
               &     $r$         &     $p$         \\

\hline

Rat: parietal cortex  & -0.992   &  0.001   \\ 
Rat: visual cortex    & -0.753   &  0.084  \\ 
Cat: visual cortex    & -0.869   &  0.011   \\
Monkey: frontal cortex  &  -0.889  &  0.044  \\
Monkey: visual cortex   &  -0.927  &  0.008  \\
Monkey: sensorimotor crtx  &  -0.995  &  0.000  \\
Human: frontal cortex      &  -0.642 (0.049) &  0.063 (0.908)    \\
Human: visual cortex       &  -0.968 (-0.968) &  0.000 (0.002)   \\
Human: temporal cortex     &  -0.659 (-0.872) &  0.054  (0.005)    \\

\hline

\hline
\end{tabular}
\end{center}
Values in the brackets refer to $r$ and $p$ without the prenatal
data points.

\end{table}

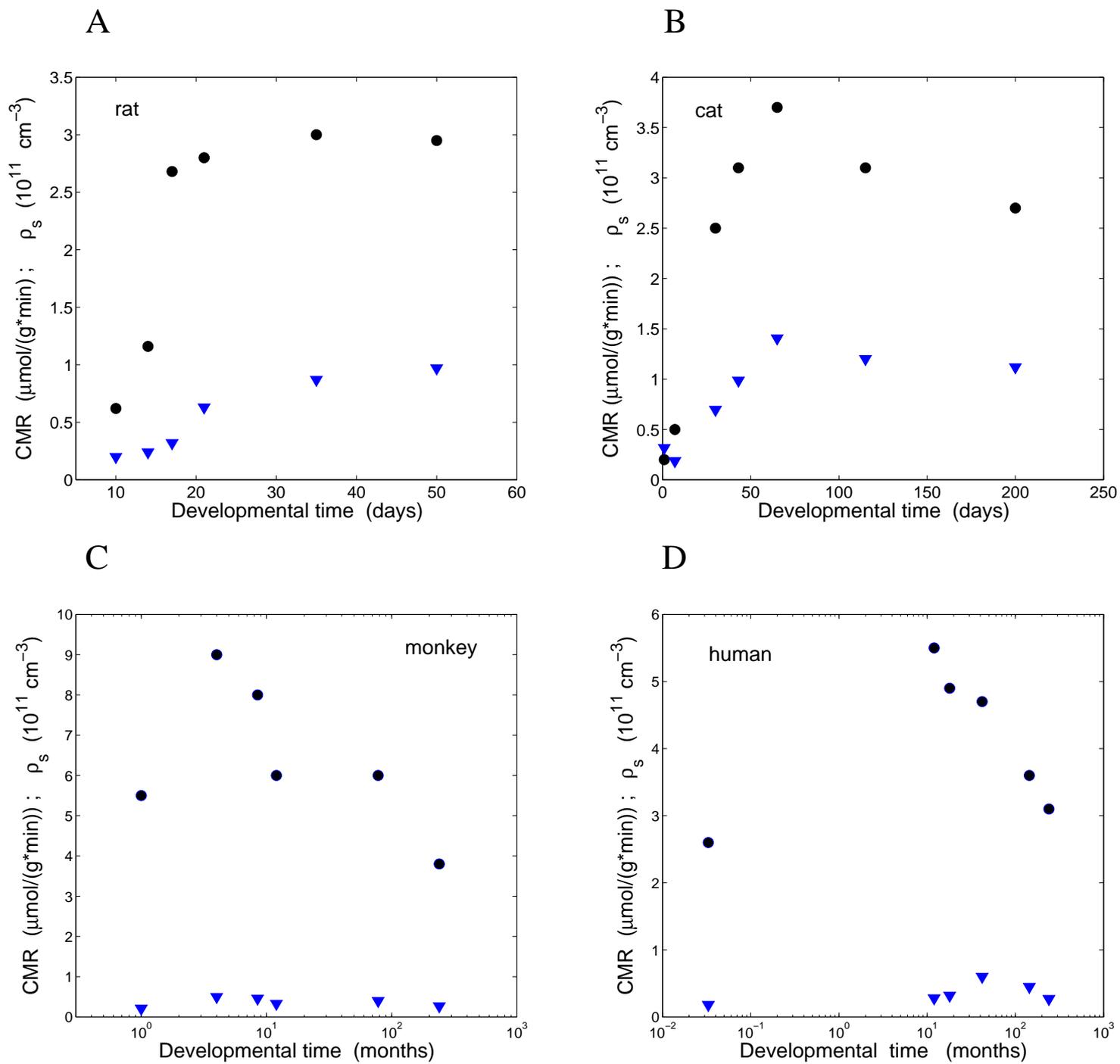
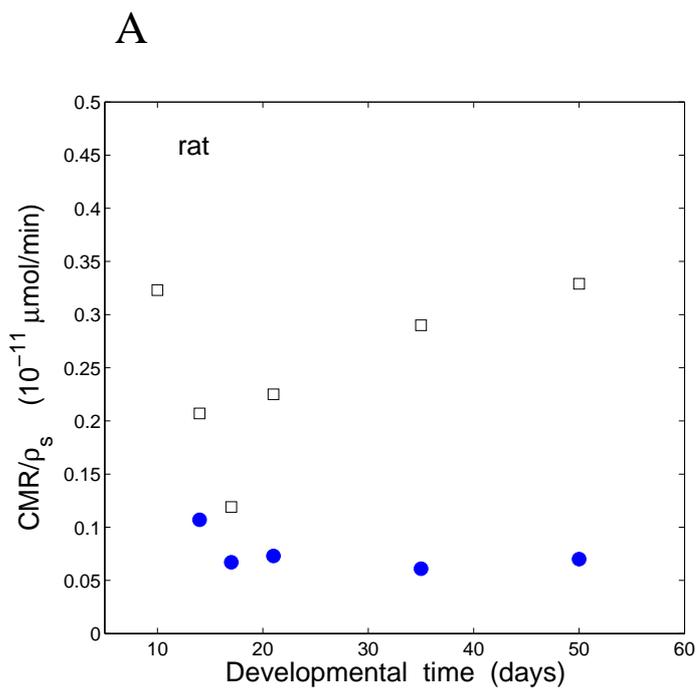
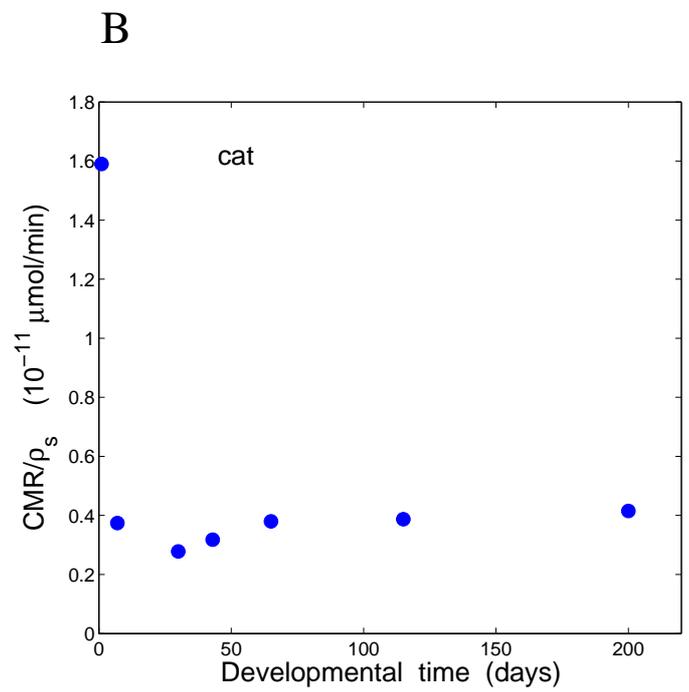
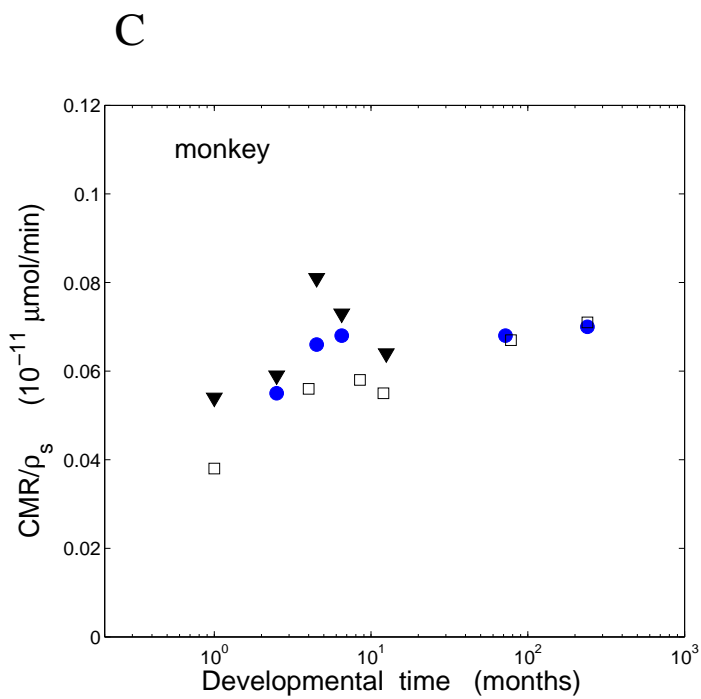
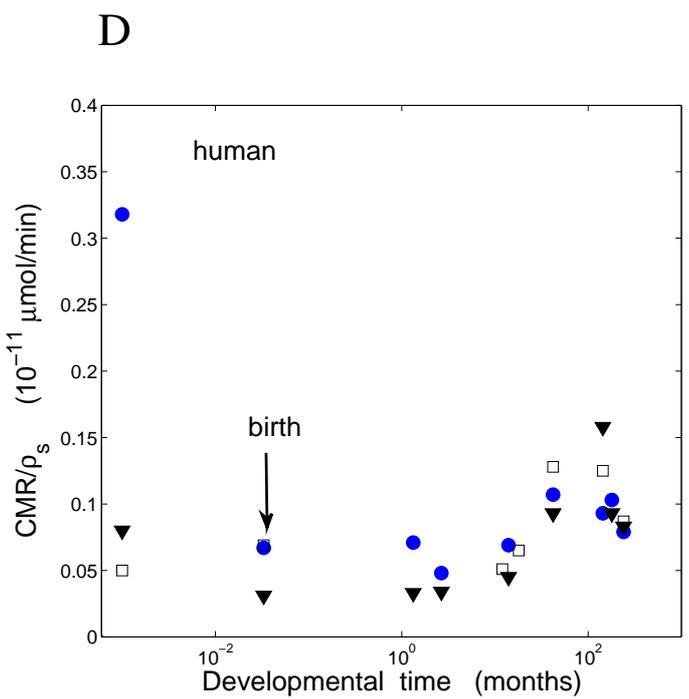
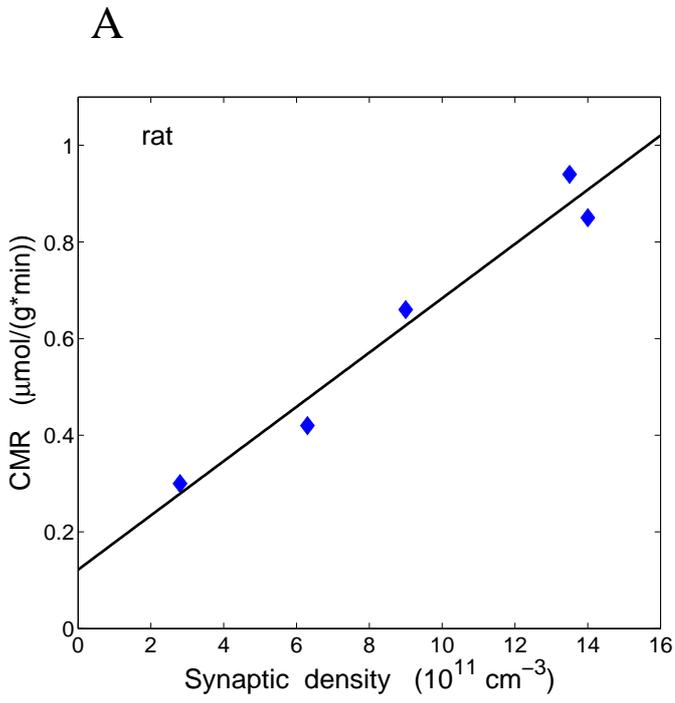
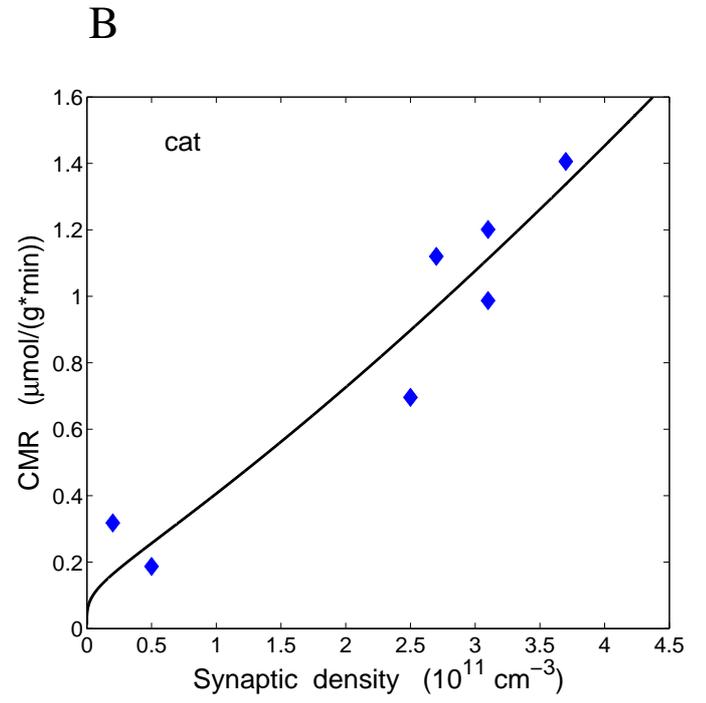
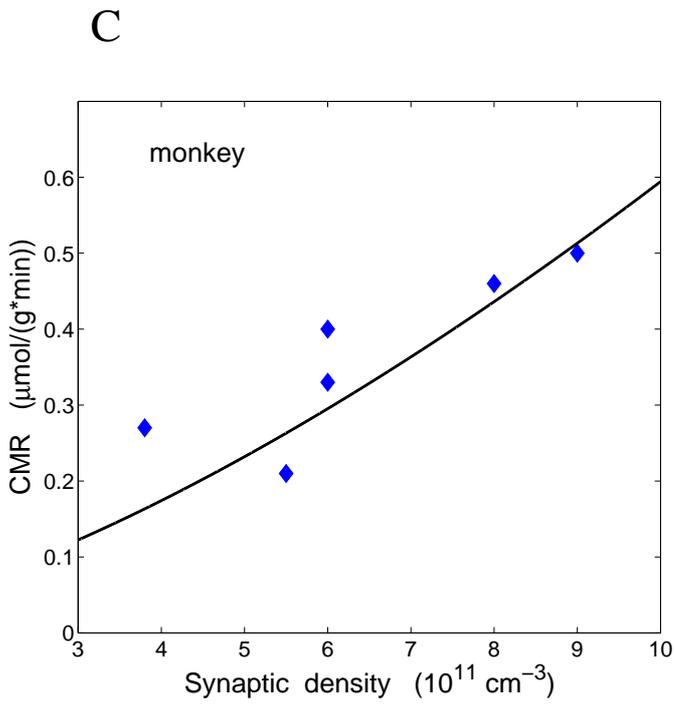
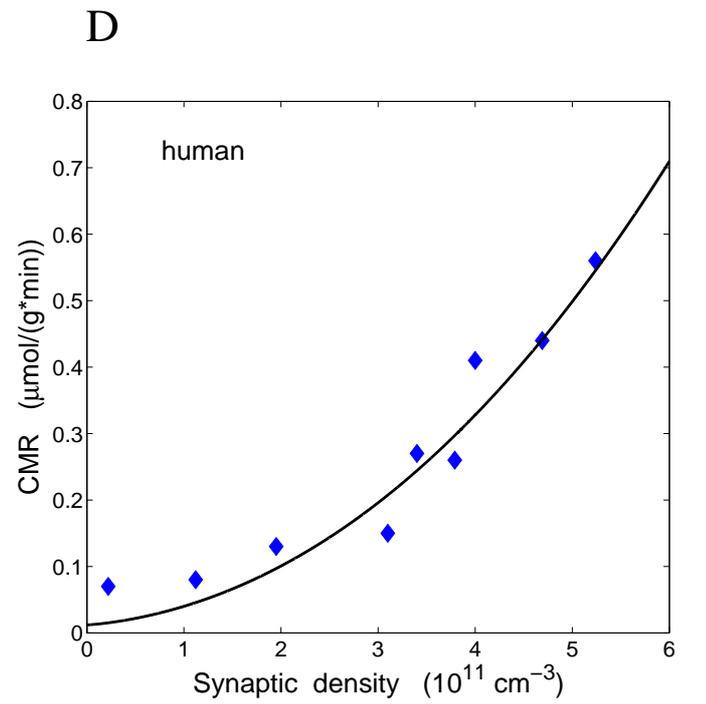
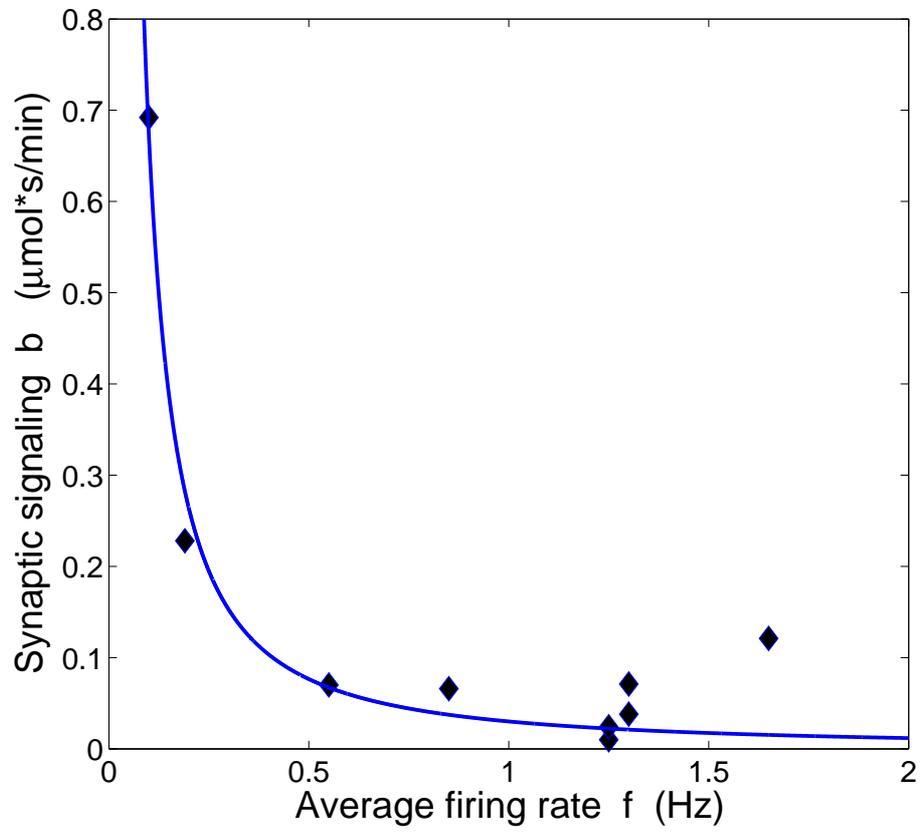
\end{document}